\documentclass[letterpaper, 10 pt, journal, twoside]{IEEEtran}
\usepackage{cite}
\usepackage{amsmath,amssymb,amsfonts}
\usepackage{algorithmic}
\usepackage{graphicx}
\usepackage{textcomp}

\usepackage[latin1]{inputenc}
\usepackage[english]{babel}
\usepackage{graphicx}
\usepackage{float}
\usepackage{psfrag}
\usepackage{amssymb}
\usepackage{amsmath}
\usepackage{amscd}
\usepackage{eucal}
\usepackage{color}
\usepackage{caption}
\usepackage{bm}
\usepackage{epsf}

\newtheorem{cor}{Corollary}
\newtheorem{prop}{Proposition}[section]
\input{epsf}
\begin{document}
\renewcommand{\figurename}{Figure}

\title{Primary Control Effort under Fluctuating Power Generation in Realistic High-Voltage Power Networks\\
}
\author{{Melvyn Tyloo}
{and Philippe Jacquod}
\IEEEmembership{Member, IEEE}
\thanks{This works was supported by the Swiss National Science Foundation under grant N$^\circ$ 200020\_182050}
\thanks{M.~Tyloo is with the School of Engineering, University of Applied Sciences of Western Switzerland HES-SO, CH-1951 Sion, Switzerland
        {\tt\small melvyn.tyloo@gmail.com}}%
\thanks{Ph.~Jacquod is with the School of Engineering, University of Applied Sciences of Western Switzerland HES-SO, CH-1951 Sion, Switzerland, and the Department of Quantum Matter Physics, University of Geneva, CH-1211 Geneva, Switzerland.}%
}





\maketitle
\thispagestyle{empty}
\begin{abstract}
Many recent works in control of electric power systems have investigated their synchronization through global 
performance metrics under external disturbances. The approach is motivated by fundamental changes in the operation of power
grids, in particular by the substitution of conventional power plants with new renewable sources
of electrical energy. This substitution will simultaneously increase fluctuations in power generation and reduce the available mechanical 
inertia. It is crucial to understand how strongly these two evolutions will impact grid stability. With very few, mostly
numerical exceptions, earlier works on performance metrics
had to rely on unrealistic assumptions of grid homogeneity. Here we show that a modified 
spectral decomposition can tackle that issue in inhomogeneous power grids in cases where disturbances occur on time scales
that are long compared to the intrinsic time scales of the grid. We find in particular that 
the magnitude of the transient excursion generated by disturbances with long characteristic times does not depend on inertia.
For continental-size, high-voltage power grids, this corresponds to power fluctuations that are correlated on time scales of few seconds or more. 
We conclude that power fluctuations arising from new renewables will not require per se the deployment of additional rotational inertia. 
We numerically illustrate our results on the IEEE 118-Bus test case and a model of the synchronous grid of continental Europe.
\end{abstract}


\section{Introduction}
\IEEEPARstart{T}{he} penetration of 
new renewable sources of electrical energy is currently increasing in most electric power grids worldwide, as more and more traditional power plants are phased out. 
A major concern is obviously that this substitution reduces the available inertia while it simultaneously induces
 larger fluctuations in power generation~\cite{Ulb14}. Both changes may jeopardize grid stability, either individually or taken together.
A key issue is accordingly to evaluate how much  power grids need to be adapted  
to their resulting new modes of operation -- for instance through line extensions or deployment of resources providing ancillary services~\cite{AnnualEnOutlook}. 
To ensure the stability of the grid and the safety of power supply, it is important to clarify the role of the generators dynamical parameters that will be affected by this transition, namely the  
rotational 
inertia and the frequency damping / droop control. Both are going to be globally reduced, moreover their geographical distribution will be modified. To try to identify
which of these dynamical parameters are most crucial, where they should be primarily deployed,  as well as when and in what operating conditions they are most needed, analytical and 
numerical works have investigated the robustness of electric power grids under external disturbances. 
The response of power grids to external disturbances has been investigated through
quadratic performance metrics~\cite{Bam13,Dor14a,Teg15,Sia16,Poo17,Pag17,Gru18,Col19,Pag19b, Yan19},
eigenvalue damping ratios and frequency overshoots~\cite{Mev16}, rate of change of frequency\cite{Pag17,Guo18} or 
disturbance wave propagation~\cite{Tam18,Hae18}. Except for numerical results~\cite{Mev16,Tam18,Hae18}, 
these works considered disturbances with infinitely short time scales, such 
as white noise power fluctuations or instantaneous power injection changes~\cite{Bam13,Teg15,Sia16,Poo17,Gru18,Col19,Pag19b, Yan19}, 
or with infinitely long time scales, such as step changes in power injection~\cite{Pag17,Guo18}. All of the analytical works 
relied on one of the two homogeneity assumptions that
inertia and frequency damping are the same everywhere, or that their ratio is. 

These homogeneity assumptions 
are not representative of real power grids, where in particular, consumer nodes are inertialess but with small, albeit finite frequency damping~\cite{Ber81,Mac08}.
Often, this inconsistency  is circumvented by invoking a prior Kron reduction absorbing the inertialess nodes into an effective network.
One then measures the robustness of that reduced network, which may or may not be related to the robustness of the original one, because Kron reduction does not 
capture the dynamics of the reduced, inertialess nodes. 
To the best of our knowledge, Ref.~\cite{Pag19b} is the first work that tolerates deviations from homogeneity in an analytical calculation of a quadratic
performance metric. Its results suggest that grid robustness is crucially sensitive to 
the geographic distribution of frequency control, while inertia has to be distributed rather evenly. This conclusion has to be revisited because,
first, Ref.~\cite{Pag19b} is based on an approximate method tolerating only small deviations from homogeneity and second, the only disturbances it considers
are long power losses. 

In this manuscript, we investigate the response of power systems to colored noisy power fluctuations.
We quantify the response to these disturbances by a quadratic performance metric measuring the primary control effort
necessary to absorb the fault~\cite{Poo17}. Our analytical approach still relies on a homogeneity assumption. 
Our results however emphasize the role played by the different time scales in the problem: 
the performance metric depends on inertia only when the characteristic time scale of the disturbance is long compared to all other time scales in the system. 
In that case we conjecture, and confirm numerically, that our analytical results also apply to heterogeneous systems. 
In continental-size 
high-voltage power grids, the network time scales are shorter than few seconds~\cite{Tyl18b}. Therefore, noise fluctuating on time scales of tens of second or
more induces transients whose amplitude, duration and oscillations are largely independent of inertia, and our analytical results directly apply to most 
disturbances on realistic, inhomogeneous high voltage power grids.


The manuscript is organized as follows. Section \ref{notation} defines our mathematical notations.
Section \ref{section2} defines the power network model and gives an analytical expression for its linear response.  Section \ref{section3} introduces performance metrics and gives analytical expressions for them. Of particular interest are the short- and long-correlation time asymptotics. In Sec.~\ref{section5}, we numerically confirm our theory on both the IEEE 118-Bus test case and the PanTaGruEl model of the synchronous grid of continental Europe. We discuss time scales in such 
high-voltage power grids and show that the corresponding performance metric is given by the long noise correlation time asymptotic limit. 
Our conclusions are given in section \ref{section6}.

\section{Mathematical Notation}\label{notation}
Given a vector $\bm v\in \mathbb{R}^n$, we denote its transpose by $\bm v^\top$. We write   $\bm M = {\rm diag}\{m_i\}$ 
for the diagonal $n \times n$ matrix with $m_1,m_2,...,m_n \in \mathbb{R}$ on its diagonal. The $j$-th unit vector with a single nonzero component is 
$(\hat{e}^j)_i=\delta_{ij}$\,. The scalar product of two vectors $\bm u, \bm v\in \mathbb{R}^n$ is written $\bm u^\top\bm v$ and the scalar product of a vector with itself
is $\bm v^2=\bm v^\top\bm v$. The statistical average of a random variable $x\in\mathbb{R}$ is $\overline{x}$\,. Finally, considering a diagonal matrix $\bm M$\,, we denote 
its $p^{\rm th}$ power as $\bm M^p={\rm diag}\{m_i^p\}$\,.

\section{Power Grids and their Response to Fluctuating Power Injections}\label{section2}
\subsection{Swing dynamics near synchrony}
Transient dynamics in high-voltage power networks is commonly modelled by the swing equations which describe the dynamics of voltage angles assuming constant voltage amplitudes. In the lossless line approximation, appropriate to very high-voltages~\cite{Mac08}, 
they read
\begin{eqnarray}\label{eq:kuramoto}
m_i\,\dot{\omega}_i+d_i\,{\omega}_i=P_i-\sum_j b_{ij}\sin(\theta_i-\theta_j),
\end{eqnarray}
where each network node is labeled $i=1,...,n$ with a voltage angle $\theta_i$. Equation~\eqref{eq:kuramoto} is written 
in a rotating frame, so that the frequency $\omega_i=\dot{\theta}_i$ refers to the deviation from the rated frequency of $50$ or $60$ Hz. Each node has inertia and damping control parameters $m_i$ and $d_i$ respectively, and an active power $P_i$ that is generated ($P_i>0$) or consumed ($P_i<0$). The coupling between node $i$ and $j$ 
is given by the susceptance $b_{ij} \ge 0$ of the corresponding power line. The operational state ${\bm \theta}^{(0)}$ is a synchronous 
stationary solution to Eq.~(\ref{eq:kuramoto}). 

Equation~(\ref{eq:kuramoto}) is governed by two sets of time scales. The first set is given by the ratio between inertia and damping coefficients $\gamma_i^{-1}\equiv m_i/d_i$. It corresponds to the local relaxation of synchronous machines. The second one is determined by the network characteristic time scales $d_i/\lambda_\alpha$ 
given by damping coefficients and the eigenvalues of the weighted Laplacian, see Eq.~(\ref{eq:laplacian}) below. 
Depending on these two sets of time scales, perturbations are locally damped or spread across the network. In a synthetic synchronous grid of continental Europe
with constant damping and inertia corresponding to the average of their true values, 
it has been found that
all modes are underdamped and propagate through the whole system with $d/\lambda_\alpha < \gamma^{-1}$, $\forall \alpha$~\cite{Tyl18b}.

We next investigate the response of the system to a time-dependent disturbance ${\bm{P}}(t) = {\bm{P}} + \delta {\bm{P}}(t)$ acting on 
the operational state ${\bm \theta}^{(0)}$, following which angles become time-dependent, 
${\bm{\theta}}(t) = {\bm{\theta}}^{(0)} + \delta {\bm{\theta}}(t)$. The small-signal response is governed by dynamical equations obtained after linearizing Eq.~(\ref{eq:kuramoto}) about 
$\bm{\theta}^{(0)} $,
\begin{align}\label{eq:kuramoto_lin}
{\bm M}\,\dot{\bm \omega} + {\bm D}\, {\bm \omega} &= \delta {\bm P}(t) - {\mathbb L}({\bm\theta}^{(0)}) \, \delta {\bm \theta} \, ,
\end{align}
where we introduced inertia and damping matrices, ${\bm M}={\mathrm{diag}}\{m_i\}$ and ${\bm D}={\mathrm{diag}}\{d_i\}$ and the weighted Laplacian matrix ${\mathbb L}(\{ \theta_i^{(0)} \})$ with matrix elements
\begin{equation}\label{eq:laplacian}
{\mathbb L}_{ij} = 
\left\{ 
\begin{array}{cc}
-b_{ij} \cos(\theta_i^{(0)} - \theta_j^{(0)}) \, , & i \ne j \, , \\
\sum_k b_{ik} \cos(\theta_i^{(0)} - \theta_k^{(0)}) \, , & i=j \, .
\end{array}
\right.
\end{equation}
This Laplacian is minus the stability matrix of the linearized dynamics, and 
since we consider a stable synchronous state, it
is positive semidefinite, with a single vanishing eigenvalue $\lambda_1=0$ with eigenvector ${\bm u}_1=(1,1,...1)/\sqrt{n}$. All other eigenvalues are positive,  
$\lambda_\alpha>0$, $\alpha=2,3,...n$. 
Equation~(\ref{eq:kuramoto_lin}) can be integrated via a spectral decomposition provided either: (i) both $\bm M$ and $\bm D$ commute with $\mathbb{L}$, then Eq.~(\ref{eq:kuramoto_lin}) can be integrated in the eigenspace of $\mathbb{L}$, or (ii)  ${{\bm M}^{-1}{\bm D}}=\gamma\,\mathbb{I}$\,, in which case Eq.~(\ref{eq:kuramoto_lin}) can be integrated in the eigenspace of ${\bm D^{-1/2}}\,\mathbb{L}\,{\bm D^{-1/2}}$. 

\subsection{Analytical solution for constant damping-to-inertia ratio}
We consider the case (ii) above of constant  inertia-to-damping ratio, $m_i/d_i=\gamma^{-1}$ $\forall i$\,. To calculate the response of the system,  
we first perform a change of variable ${\delta\bm \varphi}={\bm D}^{1/2}\delta {\bm \theta}$
on Eq.~(\ref{eq:kuramoto_lin}). We obtain
\begin{align}\label{eq:kuramoto_lin3}
\gamma^{-1}\delta \ddot{{\bm \varphi}} + \delta \dot{{\bm \varphi}} &= {\bm D}^{-1/2}\delta {\bm P} -  {\bm D}^{-1/2}\,\mathbb{L}\,{\bm D}^{-1/2}\, \delta {\bm \varphi} \, .
\end{align}
We choose this normalization with $\bm D$, rather than with $\bm M$ as proposed in Ref.~\cite{Pag17}, because it allows us to treat
the inertialess case with $\gamma^{-1}=0$, from which we will extrapolate 
the realistic case where consumer nodes are inertialess and generator nodes have nonhomogeneous inertia. 
Equation~(\ref{eq:kuramoto_lin3}) can be solved by expanding angle deviations as $\delta{\bm \varphi}(t)=\sum_{\alpha}c_\alpha(t)\,{{\bm u}_\alpha^{D}}$,
over the eigenvectors ${\bm u}_{\alpha}^{D}$ of $\mathbb{L}^D={\bm D}^{-1/2}\,\mathbb{L}\,{\bm D}^{-1/2}$.
The matrix $\mathbb{L}^D$ is no longer Laplacian but it still has a zero-mode ${\bm u}_1^D=(\sqrt{d_1},...,\sqrt{d_n})/\sqrt{\sum_i d_i}$. Angle shifts of $\delta\bm \varphi$ along ${\bm u}_1^D$ do not modify the synchronous state because ${\bm u}_1\propto{\bm D}^{-1/2}{\bm u}_1^D$. Note also that, by orthogonality with ${\bm u}_1^D$, eigenvector components 
must satisfy $\sum_i \sqrt{d_i}u_{\alpha,i}^D=0$ for $\alpha\ge 2$.

\begin{prop}\label{prop1}
\textit{The general solution to Eq.~(\ref{eq:kuramoto_lin3}) reads}
\begin{equation}\label{eq:calpha2}
\begin{aligned}
\delta\varphi_i(t)&=  \sum_\alpha \gamma e^{\frac{-\gamma-\Gamma_{\alpha}}{2}t}\int_0^{t}e^{{\Gamma_{\alpha}}t_1}  \\
&\times \int_{0}^{t_1}[{\bm D}^{-1/2}\delta {\bm{P}}(t_2)]^\top {\bm{u}}_{\alpha}^De^{\frac{\gamma-\Gamma_{\alpha}}{2}t_2}{\rm d}t_2{\rm d}t_1 \;u_{\alpha,i}^D \;,
\end{aligned}
\end{equation}
\textit{with $\Gamma_\alpha=\sqrt{\gamma^2-4\lambda_\alpha^D\gamma}$ where $\lambda_\alpha^D$ is the eigenvalue associated with the eigenvector ${\bm u}_\alpha^D$ of $\mathbb{L}^D$.}
\end{prop}
\begin{IEEEproof}
We first expand angle deviations over 
the eigenbasis of $\mathbb{L}^D$ as $\delta\varphi_i(t)=\sum_{\alpha}c_\alpha(t)\, u_{\alpha,i}^D$. From the orthogonality of the
eigenbasis, $({\bm{u}}_\alpha^D)^\top {\bm{u}}_\beta^D=\delta_{\alpha \beta}$,  one straightforwardly rewrites Eq.~(\ref{eq:kuramoto_lin3}) as
\begin{eqnarray}\label{eq:ca}
\gamma^{-1}\ddot{c}_\alpha + \dot{c}_\alpha = ({\bm D}^{-1/2}\delta {\bm P})^\top {\bm{u}}_\alpha^D - \lambda_\alpha^D c_\alpha\; ,
\end{eqnarray}
for $\alpha=1,...,n$\,. The expansion coefficients $c_\alpha(t)$ can be read from
Eq.~(\ref{eq:calpha2}), and direct differentiation shows that they
solve this equation. This completes the proof.
\end{IEEEproof}

\section{Dynamical Parameters and Transient Excursions}\label{section3}
\subsection{Quantifying frequency excursions}\label{sec:perf}
To evaluate the global response of the system to an external disturbance, we use the following performance metric
\begin{align}\label{eq:c1}
\begin{split}
{\mathcal P}(T) &= T^{-1} \, \int_0^T \,({\bm \omega}^\top-\overline{{\bm \omega}}^\top){\bm D}({\bm \omega}-\overline{{\bm \omega}})\, {\rm d}t   \, ,
\end{split}
\end{align}
because it measures the primary control effort and therefore has a physical meaning~\cite{Poo17}. The quantity 
$\overline{{\bm \omega}}^\top=(\dot\Delta,\dot\Delta,...,\dot\Delta)$ with $\dot\Delta (t)  = \sum_i d_i\omega_i(t)/\sum_i d_i$ gives the average frequency deviation over
all nodes in the system.
Because synchronous states are defined modulo any
homogeneous angle shift, the transformation $\theta_i^{(0)} \rightarrow \theta_i^{(0)} + C$ does not change the 
synchronous state. Accordingly only phase and frequency shifts with $\sum_i \delta \theta_i(t)=0$ and $\sum_i d_i \omega_i(t)=0$ matter. This is included
in ${\mathcal P}$ by subtracting the average $\overline{{\bm \omega}}$. That ${\mathcal P}(T)$ is a performance metric is easily understood: 
low values indicate that the system absorbs the perturbation with little fluctuations, while large values
indicate a large transient excursion around the initial synchronous state. 
Using the above change of variables, $\delta\dot{{\bm{\varphi}}}={\bm D}^{1/2}{\bm \omega}=\sum_{\alpha}\dot{c}_\alpha(t)\, {\bm u}_{\alpha}^D$\,, 
Eq.~(\ref{eq:c1}) becomes,
\begin{align}\label{eq:perf2}
\begin{split}
\mathcal{P}(T)&=T^{-1} \sum_{\alpha \ge 2}\int_0^T\dot{c}_\alpha^2(t){\rm d}t \; .
\end{split}
\end{align}
This can be calculated using  the explicit expression for $c_\alpha(t)$ from Eq.~(\ref{eq:calpha2}), once a perturbation $\delta {\bm P}(t)$ is given.\\
\begin{prop}\label{prop2}
\textit{Consider a noisy disturbance acting on $N_n$ of the network nodes. The noise ensemble is Gaussian and defined by its
vanishing first moments, $\overline{\delta P_i(t)}=0$, and its second moments $\overline{\delta P_i(t)}\overline{\delta P_j(t')}=\delta_{ij}\delta P_{0i}^2\exp[-|t-t'|/\tau_0]$ with the noise correlation time $\tau_0$. The performance metric $\overline{{\mathcal P}^\infty}$ for primary control effort averaged over this noise ensemble is given by}
\begin{align}\label{eq:P2}
\begin{split}
\overline{\mathcal{P}^\infty}&=\sum_{\alpha\ge 2}\frac{\sum_{i\in N_n} \delta P_{0i}^2{u_{\alpha,i}^D}^2d_i^{-1}}{\lambda_\alpha^D\tau_0+1+\gamma^{-1}\tau_0^{-1}} \; .
\end{split}
\end{align}
\end{prop}
\begin{cor}
\textit{In the limit of short noise correlation time, $\tau_0\ll \gamma^{-1},{\lambda_\alpha^D}^{-1}$ one has}
\begin{align}
\begin{split}\overline{\mathcal{P}^\infty}&=\tau_0 {\sum_{i\in N_n} \delta P_{0i}^2  \, ( 1/m_i - 1 / \sum_j m_j )}\; .\label{eq:P2s}
\end{split}
\end{align}
\end{cor}
\begin{cor}
\textit{In the opposite asymptotic limit, $\tau_0\gg \gamma^{-1},{\lambda_\alpha^D}^{-1}$ one has}
\begin{align}
\begin{split}
\overline{\mathcal{P}^\infty}&=\tau_0^{-1}\sum_{\alpha\ge 2}\frac{\sum_{i\in N_n} \delta P_{0i}^2{u_{\alpha,i}^D}^2d_i^{-1}}{{\lambda_\alpha^D}}  \; .\label{eq:P2l}
\end{split}
\end{align}
\end{cor}
\begin{IEEEproof} Inserting the time derivative of Eq.~\eqref{eq:calpha2} into Eq.~(\ref{eq:perf2}) and taking the average over the noise ensemble 
with $\overline{\delta P_i(t)}\overline{\delta P_j(t')}=\delta_{ij}\delta P_{0i}^2\exp[-|t-t'|/\tau_0]$  
gives Eq.~\eqref{eq:P2}, with few straightforwardly calculated exponential integrals. The two asymptotic results \eqref{eq:P2s} and \eqref{eq:P2l}
are easily obtained by a Taylor expansion, keeping only the first non-vanishing term. To obtain Eq.~\eqref{eq:P2s},
we also used $\sum_{\alpha\ge 2}{u_{\alpha,i}^D}^2 = \sum_{\alpha\ge 1}{u_{\alpha,i}^D}^2 -  {u_{\alpha,1}^D}^2 =1-d_i/\sum_i d_i$.\\
\end{IEEEproof}

{\textit{Remark 1: The short correlation time asymptotic of Eq.~\eqref{eq:P2s} agrees with the result of~\cite{Poo17} obtained for either single-pulsed or
averaged white-noise perturbations.}}

{\textit{Remark 2: The noise correlators are defined either as time averages, $\overline{\delta P_i(t)}\overline{\delta P_j(t')} = {\rm lim}_{\tau \rightarrow \infty} \tau^{-1}
\int_0^\tau \delta P_i(t+\tau') \delta P_j(t'+\tau') {\rm d}\tau$ or as averages over different noise sequences. 
 }}
 
{\textit{Remark 3: Finite-time corrections to Eqs.~\eqref{eq:P2}--\eqref{eq:P2l} disappear 
as ${\cal O}(1/T)$ as $T \rightarrow \infty$.
 }}\\ 

\begin{prop}\label{prop3}
Under the same assumptions as Proposition~\ref{prop2}, the variance ${\rm var}\left[{\mathcal P}(T)\right]$ of the performance metric  for primary control effort 
over the noise ensemble vanishes as $\sim T^{-1} + {\cal O}(1/T^2)$ as $T \rightarrow \infty$.
\end{prop}
The proof proceeds through direct calculation of ${\rm var}\left[{\mathcal P}(T)\right]$. It
is too long to fit in this article and here we only sketch it. 
From Eq.~\eqref{eq:perf2} one has
\begin{align}\label{eq:varP}
\begin{split}
{\rm var}\left[{\mathcal P}(T) \right]&=
T^{-2} \sum_{\alpha,\beta \ge 2}\iint_0^T \overline{\dot{c}_\alpha^2(t)\dot{c}_\beta^2(t')}{\rm d}t {\rm d}t' - \overline{\mathcal{P}(T)}^2 \, . 
\end{split}
\end{align}
From Eq.~\eqref{eq:calpha2}, 
each $\dot{c}_{\alpha,\beta}$ contains a noise term $\delta {\bm P}$. The noise average in the first term on the right-hand side of
Eq.~\eqref{eq:varP} therefore consists in pairings of four noise terms. 
There are three such contributions. The first one pairs the two $\delta {\bm P}$'s in $\dot{c}_{\alpha}^2(t)$
and the two $\delta {\bm P}$'s in $\dot{c}_{\beta}^2(t)$. This contribution is cancelled by a similar pairing in $\overline{\mathcal{P}(T)}^2$. The other two contributions 
pair $\delta {\bm P}$'s across indices $\alpha$ and $\beta$ and accordingly, they constrain the values that $t$ and $t'$ can take with respect to one another,
$|t-t'| \lesssim \tau_0$. Accordingly, the double time integral in Eq.~\eqref{eq:varP} gives a contribution $\sim T \tau_0$, instead of $\sim T^2$, resulting in 
${\rm var}\left[{\mathcal P}(T)\right] \sim T^{-1}$.

\begin{figure*}
\centering
\includegraphics[width=0.9\textwidth]{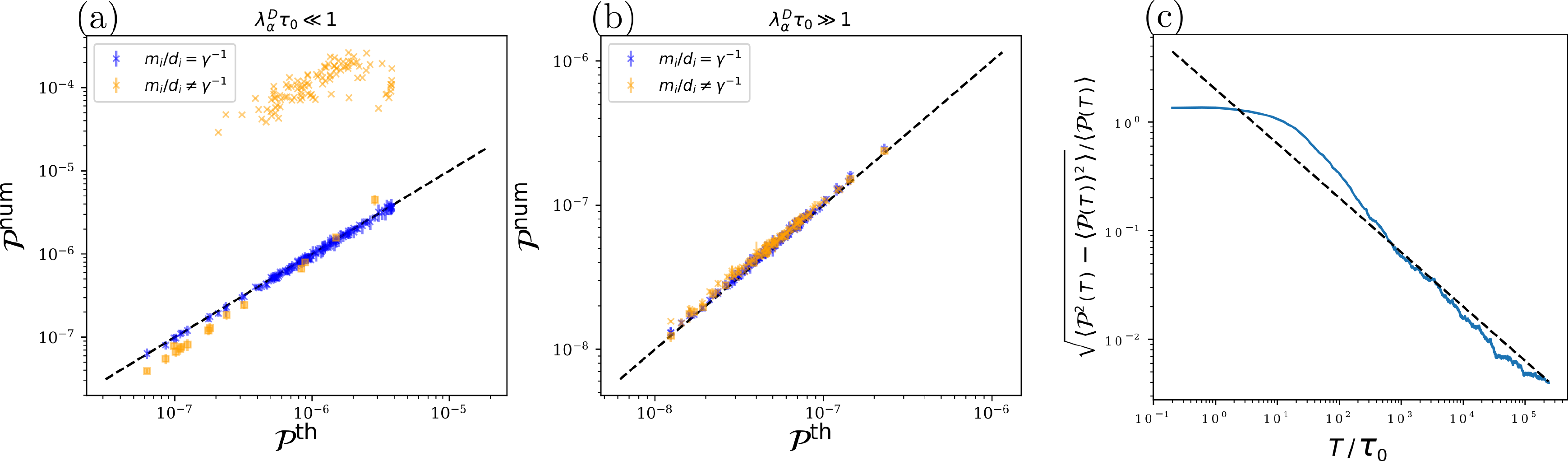}
\caption{Comparison between numerical calculations obtained by time-evolving Eq.~(\ref{eq:kuramoto}) and the theoretical result of Eq.~(\ref{eq:P2l}) for the primary control effort,
for short correlation time $\gamma\tau_0 = 4\times 10^{-3}$ (a) and long correlation time $\gamma\tau_0 = 40$ (b)
in the IEEE 118-Bus test case. Averages are made over $10$ noisy sequences and standard deviations are shown by barely visible vertical line. Blue crosses correspond to constant damping-to-inertia ratio, $d_i/m_i=\gamma=0.4s^{-1}$, while orange symbols correspond to inhomogeneous,
non-vanishing inertia on generator nodes (squares) and inertialess consumer nodes (crosses). On generation nodes one has $\gamma_i\tau_0\in [30 , 60]$\,. (c) Ratio between standard deviation and average of the primary control effort for the IEEE 118-Bus test case. For long enough $T$, the ratio scales as $T^{-1/2}$, confirming Prop.~\ref{prop3}. Averages are made over $40$ different noise sequences.}\label{fig2}
\end{figure*}

{\textit{Remark 4: Proposition~\ref{prop3} means that for specific noisy disturbances satisfying the 
assumption of Proposition~\ref{prop2} and for long enough observation times $T \gg \tau_0$, 
$\mathcal{P}^\infty = \overline{\mathcal{P}^\infty} + {\cal O}(T^{-1/2})$.  The statistical average is therefore representative of specific noise disturbances for 
sufficiently long observation time. The validity of Proposition~\ref{prop3} is illustrated numerically in Fig.~\ref{fig2} (c).}}

The two asymptotic limits of large and small $\tau_0$ are particularly interesting as they shed light on the influence of dynamical parameters,
in particular on the interplay between local disturbance absorption by inertia and long-range propagation through low-lying network modes.  
First, in the short correlation time limit given by Eq.~(\ref{eq:P2s}),  $\overline{\mathcal{P}^\infty}$ explicitly depends on inertia but not on the 
coupling network. This reflects the fact that, in the white-noise limit, the perturbation remains local and is easily absorbed, if 
there is enough inertia. Second, 
Eq.~(\ref{eq:P2l}) shows that, in the long correlation time limit, $\overline{\mathcal{P}^\infty}$ does not depend on 
inertia. This suggests that changing inertia in any direction
will not change $\overline{\mathcal{P}^\infty}$ in the limit of long noise correlation time. This is a conjecture since Eq.~(\ref{eq:P2l}) has been derived 
under the assumption of constant damping-to-inertia ratio, $\gamma=d_i/m_i$. Below we numerically confirm this conjecture.
Simultaneously, Eq.~(\ref{eq:P2l}) also shows that, in the long correlation time limit, $\overline{\mathcal{P}^\infty}$
is determined by the structure of the coupling network, with the modes with smallest eigenvalues having the largest influence. Those modes 
are extended over the whole network in large power grids, as is illustrated in Fig.~\ref{fig3} (b). Accordingly, in the limit of long noise correlation time, 
the disturbance is able to propagate over large distances in the network, and inertia has little influence on this large-scale propagation. 

{\textit{Remark 5: It is important to realize that
fluctuations of renewable energy sources occur on time scales that are large compared to the intrinsic time scales of the power system~\cite{Mei11}. As a matter of fact, time scales in the synchronous grid of continental Europe have been found to 
satisfy $\gamma^{-1} \simeq 2.5 s$ and  ${\lambda_\alpha^D}^{-1} \lesssim 0.5 s$~\cite{Tyl18b}.
Provided the above conjecture is corroborated, Eqs.~\eqref{eq:P2}--\eqref{eq:P2l} suggest that fluctuations from new renewables excite network modes and
are efficiently absorbed by optimizing the distribution of damping with little regard for inertia. This conjecture is numerically confirmed below.}

{\textit{Remark 6: Similar conclusions as in Eqs.~\eqref{eq:P2}--\eqref{eq:P2l} regarding local inertia absorption vs. large-scale mode propagation
are obtained in the case of step disturbances corresponding to sudden power losses, as a function of their duration $\tau_0$.}

{\textit{Remark 7: The lossless line approximation used in this paper does not account for ohmic dissipation. We expect that the latter enhances mode damping 
and accordingly undermines disturbance propagation in the case of noise with long correlation time, but that it affects only marginally our result for short correlation time. 
Investigations beyond the lossless line approximation would be very welcome but lie beyond the scope of the present paper.}

\section{Numerical Simulations}\label{section5}
\subsection{Dynamical parameters for simulations}\label{section5a}
We consider two different cases: (i) cases with homogeneous damping-to-inertia ratio $d_i/m_i=\gamma$; (ii) realistic heterogeneous cases, using the nonlinear swing dynamics of Eq.~(\ref{eq:kuramoto}).
In the homogeneous case, we used $d_i = \alpha |P_i^{(0)}|/\omega_0$, where $P_i^{(0)}$ is the produced/consumed power at nominal frequency $\omega_0$, 
and set $m_i=\gamma^{-1}d_i$. 
In the heterogeneous case, the inertia parameter vanishes on consumer nodes and is $m_i = 2H_i |P_i^{(0)}|/\omega_0$ on generator nodes, 
where $H_i$ depends on the type of generator~\cite{Mac08}.
Damping is given by Eq.~(5.24) and Table 4.3 in Ref.~\cite{Mac08} for generators and by 
$d_i = \alpha |P_i^{(0)}|/\omega_0$ for consumer nodes. In all 
cases we use $\omega_0=2\pi\times 50Hz$. For the IEEE-118 Bus test case discussed in Section \ref{sec:ieee118}, $\alpha=1.5$ and $H_i=5s$. For the PanTaGruEl European model discussed in Section \ref{sec:panta}), $\alpha=1.5$ and $H_i$ varies according to the generator type as described in Ref.~\cite{Pag19c}.

\subsection{IEEE 118-Bus test case}\label{sec:ieee118}
The main prediction from Eqs.~\eqref{eq:P2}-\eqref{eq:P2l} is that, for noise correlation $\tau_0$ that is longer than the 
other characteristic time scales in the system, the performance metric $\overline{\mathcal{P}^\infty}$ does not depend on  inertia parameters. This was conjectured from 
Eq.~\eqref{eq:P2l} where inertia does not appear. 
Fig.~\ref{fig2} shows performance metric obtained from individual noisy disturbance of a single node, repeating the operation for all nodes in the IEEE 118-Bus test case. 
As disturbance, we take Gaussian noise with the same first two moments as in Proposition~\ref{prop2}.
First, the primary control effort $\overline{\mathcal{P}^\infty}$ is calculated for a constant damping-to-inertia ratio (blue crosses), then for a distribution of dynamical parameters where generator nodes have inertia while consumer nodes do not (orange squares and crosses). As predicted by Eq.~(\ref{eq:P2l}), for long correlation time of the noise [panel (b), $\gamma \tau_0=40$], 
the two distributions of dynamical parameters give the same $\overline{\mathcal{P}^\infty}$, corroborating our conjecture that it does not depend on inertia. We have found (but do not show) that 
the performance metric only depends on the damping distribution in that case. 

For short noise correlation time, on the other hand, 
Eq.~(\ref{eq:P2s}) explicitly depends on the damping-to-inertia ratio $d_i/m_i=\gamma$\,, and we expect that the numerical data will differ from the theoretical 
prediction once this ratio is no longer constant. This is confirmed in 
Fig.~\ref{fig2} [panel (a), $\gamma \tau_0=4\times 10^{-3}$] where with $d_i/m_i=\gamma$, numerical data fall on the theory (blue crosses). However, once 
$d_i/m_i$ is no longer constant, numerical data and theoretical prediction differ significantly (orange symbols). 
Quite interestingly, we found that for noisy perturbations on generator nodes
with inertia, the theory still gives a remarkably accurate estimate for the primary control effort $\mathcal{P}^\infty$. An understanding of this remarkable agreement would be highly welcome, particularly since it could justify dynamic performance analysis on Kron reduced networks. 
%

\subsection{Time scales in high-voltage electric power grids}

We have shown that the primary control effort against fluctuating disturbances in the form of colored noise 
behaves very differently depending on the position of the noise correlation time relative to the characteristic time scales in the system. 
Furthermore the primary control effort is captured by our theory even for inhomogeneous dynamical parameters, when the noise correlation time is long enough. 
It is therefore desirable to identify 
what regime applies to a realistic high-voltage power grid subjected to fluctuating sources of power, in particular those generated by 
new renewable sources of energy. 
To that end, we consider in the next paragraph a realistic model of the synchronous grid of continental Europe~\cite{Tyl18b,Pag19c},
with the following time scales 
\begin{subequations}\label{eq:time}
\begin{align}
{\lambda_\alpha^D}^{-1}&< 0.5s\;, \; {\text{for $\alpha=2,...,n$}\,,}\\
\gamma^{-1}&=\frac{\langle m_i \rangle}{\langle d_i \rangle}= 2.5s\, ,
\end{align}
\end{subequations}
where $\langle \;\ldots\; \rangle$ means that we take the average over all nodes in the grid. It is commonly accepted that 
power fluctuations from renewable energy sources such as wind turbines or photovoltaic panels
fluctuate on time scales that are larger than both time scales in Eq.~(\ref{eq:time})~\cite{Mei11}. Therefore, the asymptotic limit 
of large noise correlation time, corresponding to Eq.~(\ref{eq:P2l}) applies, and we expect that the primary control effort as measured by 
Eq.~\eqref{eq:c1} is influenced only by damping, and not by inertia. Numerical results to be presented in the next paragraph corroborate this expectation.

As a side-remark we note that when the perturbation corresponds to the sudden disconnection of a power generator, 
controls usually try to reconnect the bus several times quickly after the fault (typically within few AC cycles). 
The typical time scale for such a perturbation is then less than several if not all system's time scales, and inertia obviously matters to absorb such sudden faults, as it is predicted by Eq.~(\ref{eq:P2s}). 

\subsection{The PanTaGruEl European model}\label{sec:panta}
To further illustrate the influence of dynamical parameters on the primary control effort, 
we numerically compute Eq.~(\ref{eq:c1}) on the large-scale PanTaGruEl model of the European high-voltage transmission grid~\cite{Pag19c}.
The model is shown on Fig.~\ref{fig3} (b). It has 3809 nodes and 4944 lines. More details can be found in Ref.~\cite{Pag19c}. 
We considered three different cases, (i) a homogeneous situation that corresponds to today's grid in terms of its global amount of inertia, with $\gamma\tau_0=4$ and $\gamma=d_i/m_i$ constant, 
(ii) a homogeneous situation where the inertia is reduced by a factor $10$\,, i.e. $\gamma\tau_0=40$, and $\gamma=d_i/m_i$ constant, 
and (iii) a realistic situation with inhomogeneous damping parameters and where inertia vanishes on consumer nodes and is inhomogeneous on production nodes
as described in Section~\ref{section5a} and Ref.~\cite{Pag19c}, with 
$\tau_0$ larger than all other time scales in the system. 
Remarkably, Fig.~\ref{fig3} (a) shows that the primary control effort for all cases is well predicted by Eq.~(\ref{eq:P2l}). This confirms our main finding
that, for fluctuations with a correlation time longer than any other characteristic  time scale in the system, inertia does not affect the primary control effort, Eq.~(\ref{eq:P2l}).
Quite surprisingly, an overall reduction of the total available inertia by a factor of 10 does not affect the primary control 
effort, Eq.~\eqref{eq:c1}.
%
\begin{figure*}
\centering
\includegraphics[width=0.95\textwidth]{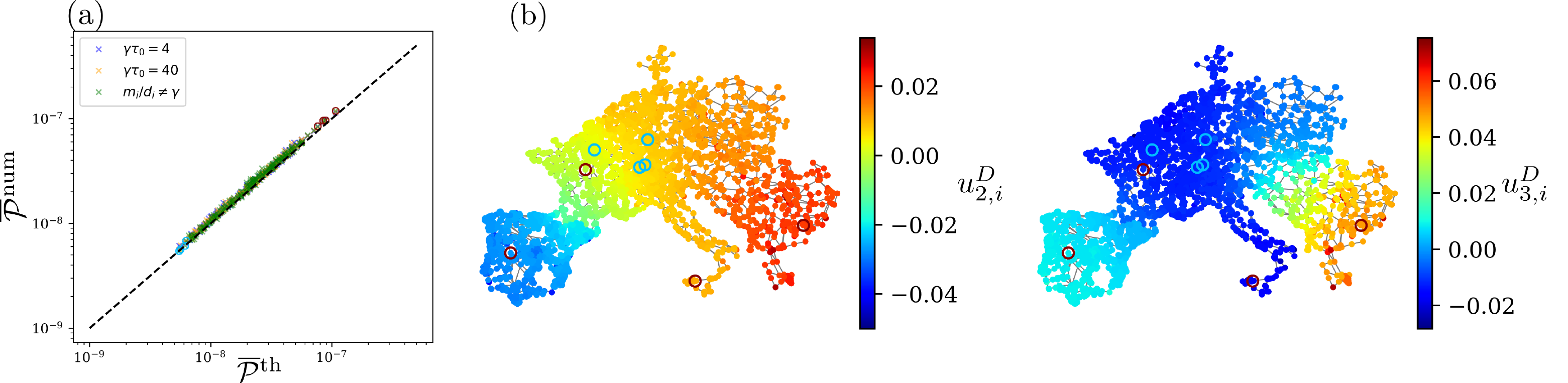}
\caption{(a) Comparison between numerical calculations obtained by time-evolving Eq.~(\ref{eq:kuramoto}) and the theoretical result of Eq.~(\ref{eq:P2l}) for the PanTaGruEl model of the synchronous grid
of continental Europe~\cite{Pag19c} shown on panel (b). 
Three cases are considered. The first two are grids with $\gamma=d_i/m_i$ constant,
with today's average inertia (orange crosses) and an inertia reduced by a factor of 10 (blue crosses). The third case corresponds to a realistic situations 
as discussed in the text, with  $\gamma_i\tau_0\in [ 20,1600 ]$ on generation nodes.
In all cases, $\tau_0$ is the longest time scale, 
consequently, the inertia-independent theoretical prediction of Eq.~\eqref{eq:P2l} accurately captures all numerical data. Averages are made over $10$ noisy sequences. (b) Network eigenmodes of $\mathbb{L}^D$
with the first two non-vanishing eigenvalues. These slow modes are extended over the whole network, with higher amplitudes on peripheral nodes. 
Disturbances on the four buses highlighted in light blue correspond to the smallest primary control effort. These buses 
lay in the center of the network where the slow modes have small amplitudes.
Disturbances on the four buses in dark red, on the other hand, have largest primary control effort. They are located at the periphery of the network
where the slow modes have large amplitudes~\cite{Tyl18b}. This shows that large primary control effort for noise with large correlation time correspond
to excitations of slow network modes, which in their turn propagate the disturbance over large distances in the network.}\label{fig3}
\end{figure*}

\section{Conclusion}\label{section6}
With the ongoing energy transition resulting in strongly increased penetrations of new renewable sources of electrical energy, a question of crucial importance is how
grid stability will evolve, given the resulting reduction of globally available rotational inertia and enhanced power fluctuations. We have shown that reduced inertia may pose 
problems only for perturbation occurring/fluctuating on very short time scales, shorter than all other characteristic time scales in the system. In continental-size transmission grids, 
these time scales are shorter than few seconds, consequently, power fluctuations from new renewables will not affect grid stability per se. 

Inertia is of course important to absorb 
sudden faults occurring on very short time scales such as line faults or disconnection/reconnection of large power plants and so forth. Simultaneously, our result
of Eq.~\eqref{eq:P2s} indicates that the resulting primary control effort is independent of the grid topology. Accordingly,
optimal inertia distribution needs to follow the distribution of potential faults, for instance being larger in regions with higher density of generators. A similar conclusion was
drawn in Refs.~\cite{Poo17} and \cite{Pag19b}. 




\begin{thebibliography}{10}
\providecommand{\url}[1]{#1}
\csname url@samestyle\endcsname
\providecommand{\newblock}{\relax}
\providecommand{\bibinfo}[2]{#2}
\providecommand{\BIBentrySTDinterwordspacing}{\spaceskip=0pt\relax}
\providecommand{\BIBentryALTinterwordstretchfactor}{4}
\providecommand{\BIBentryALTinterwordspacing}{\spaceskip=\fontdimen2\font plus
\BIBentryALTinterwordstretchfactor\fontdimen3\font minus
  \fontdimen4\font\relax}
\providecommand{\BIBforeignlanguage}[2]{{%
\expandafter\ifx\csname l@#1\endcsname\relax
\typeout{** WARNING: IEEEtran.bst: No hyphenation pattern has been}%
\typeout{** loaded for the language `#1'. Using the pattern for}%
\typeout{** the default language instead.}%
\else
\language=\csname l@#1\endcsname
\fi
#2}}
\providecommand{\BIBdecl}{\relax}
\BIBdecl

\bibitem{Ulb14}
A.~Ulbig, T.~S. Borsche, and G.~Andersson, ``Impact of low rotational inertia
  on power system stability and operation,'' \emph{IFAC Proceedings Volumes},
  vol.~47, no.~3, pp. 7290--7297, 2014.

\bibitem{AnnualEnOutlook}
``Power systems of the future: The case for energy storage, distributed
  generation, and microgrids,'' Tech. Rep., {\relax IEEE Smart Grid, Tech. Rep.
  Nov. 2012}.

\bibitem{Bam13}
B.~{Bamieh} and D.~F. {Gayme}, ``The price of synchrony: Resistive losses due
  to phase synchronization in power networks,'' \emph{\textit{American Control
  Conference}}, pp. 5815--5820, June 2013.

\bibitem{Dor14a}
F.~D\"orfler, M.~Jovanovic, M.~Chertkov, and F.~Bullo, ``Sparsity-promoting
  optimal wide-area control of power networks,'' \emph{IEEE Trans. Power
  Syst.}, vol.~29, no.~5, pp. 2281--2291, 2014.

\bibitem{Teg15}
E.~Tegling, B.~Bamieh, and D.~F. Gayme, ``The price of synchrony: Evaluating
  the resistive losses in synchronizing power networks,'' \emph{{IEEE} Trans.
  Control Net. Syst.}, vol.~2, no.~3, pp. 254--266, 2015.

\bibitem{Sia16}
M.~{Siami} and N.~{Motee}, ``Fundamental limits and tradeoffs on disturbance
  propagation in linear dynamical networks,'' \emph{IEEE Transactions on
  Automatic Control}, vol.~61, no.~12, pp. 4055--4062, Dec 2016.

\bibitem{Poo17}
B.~K. Poolla, S.~Bolognani, and F.~D\"orfler, ``Optimal {Placement} of
  {Virtual} {Inertia} in {Power} {Grids},'' \emph{IEEE Transaction Automatic
  Control}, vol.~62, no.~12, pp. 6209--6220, 2017.

\bibitem{Pag17}
F.~Paganini and E.~Mallada, ``Global performance metrics for synchronization of
  heterogeneously rated power systems: The role of machine models and
  inertia,'' \emph{\textit{55th Annual Allerton Conference on Communication,
  Control, and Computing}}, pp. 324--331, 2017.

\bibitem{Gru18}
T.~W. {Grunberg} and D.~F. {Gayme}, ``Performance measures for linear
  oscillator networks over arbitrary graphs,'' \emph{IEEE Transactions on
  Control of Network Systems}, vol.~5, no.~1, pp. 456--468, March 2018.

\bibitem{Col19}
T.~{Coletta} and P.~{Jacquod}, ``Performance measures in electric power
  networks under line contingencies,'' \emph{IEEE Transactions on Control of
  Network Systems}, vol.~7, pp. 221--231, 2020.

\bibitem{Pag19b}
L.~{Pagnier} and P.~{Jacquod}, ``Optimal placement of inertia and primary
  control: A matrix perturbation theory approach,'' \emph{IEEE Access}, vol.~7,
  pp. 145\,889--145\,900, 2019.

\bibitem{Yan19}
Y.~Jiang, R.~Pates, and E.~Mallada, ``Dynamic droop control in low-inertia
  power systems,'' 2019.

\bibitem{Mev16}
A.~Me{\v{s}}anovi{\'c}, U.~M{\"u}nz, and C.~Heyde, ``Comparison of
  ${H}_\infty$, ${H}_2$, and pole optimization for power system oscillation
  damping with remote renewable generation,'' \emph{IFAC-PapersOnLine},
  vol.~49, no.~27, pp. 103--108, 2016.

\bibitem{Guo18}
L.~{Guo}, C.~{Zhao}, and S.~H. {Low}, ``Graph laplacian spectrum and primary
  frequency regulation,'' \emph{\textit{IEEE Conference on Decision and
  Control}}, pp. 158--165, 2018.

\bibitem{Tam18}
S.~Tamrakar, M.~Conrath, and S.~Kettemann, ``Propagation of disturbances in ac
  electricity grids,'' \emph{Scientific reports}, vol.~8, no.~1, p. 6459, 2018.

\bibitem{Hae18}
H.~Haehne, K.~Schmietendorf, S.~Tamrakar, J.~Peinke, and S.~Kettemann,
  ``Propagation of wind-power-induced fluctuations in power grids,''
  \emph{Phys. Rev. E}, vol.~99, no.~5, p. 050301, 2019.

\bibitem{Ber81}
A.~R. Bergen and D.~J. Hill, ``A structure preserving model for power system
  stability analysis,'' \emph{IEEE Transaction on Power Appartus and Systems},
  vol. PAS-100, p.~25, 1981.

\bibitem{Mac08}
J.~Machowski, J.~W. Bialek, and J.~R. Bumby, \emph{Power System Dynamics},
  2nd~ed.\hskip 1em plus 0.5em minus 0.4em\relax Chichester, U.K: Wiley, 2008.

\bibitem{Tyl18b}
M.~Tyloo, L.~Pagnier, and P.~Jacquod, ``The key player problem in complex
  oscillator networks and electric power grids: Resistance centralities
  identify local vulnerabilities,'' \emph{Science Advances}, vol.~5, no.~11,
  2019.

\bibitem{Mei11}
A.~{von Meier}, ``Integration of renewable generation in california:
  Coordination challenges in time and space,'' in \emph{11th International
  Conference on Electrical Power Quality and Utilisation}, Oct 2011, pp. 1--6.

\bibitem{Pag19c}
\BIBentryALTinterwordspacing
L.~Pagnier and P.~Jacquod, ``{PanTaGruEl - a pan-European transmission grid and
  electricity generation model},'' Dec. 2019. [Online]. Available:
  \url{https://doi.org/10.5281/zenodo.2642175}
\BIBentrySTDinterwordspacing

\end{thebibliography}
\end{document}